\documentclass[useAMS,usenatbib]{mn2e}
\usepackage{graphicx}
\title[Parent Stars of Extrasolar Planets. XV]{Parent Stars of Extrasolar Planets. XV. Host Star Rotation Revisited with {\it Kepler} Data}
\author[G.\ Gonzalez]{Guillermo Gonzalez$^{1}$\\
$^{1}$Department of Physics and Astronomy, Ball State University, Muncie, IN 47306, USA\\
}

\begin{document}

\date{Accepted ??. Received ??; in original form ??}

\pagerange{\pageref{firstpage}--\pageref{lastpage}} \pubyear{??}

\maketitle

\label{firstpage}

\begin{abstract}
We employed published rotation periods of {\it Kepler} field stars to test whether stars hosting planets tend to rotate more slowly than stars without known planets. Spectroscopic vsini observations of nearby stars with planets have indicated that they tend to have smaller visni values. We employ data for {\it Kepler} Objects of Interest (KOIs) from the first 16 quarters of its original mission; stellar parameters are based on the analysis of the first 17 quarters. We confirm that KOI stars rotate more slowly with much greater confidence than we had previously found for nearby stars with planets. Furthermore, we find that stars with planets of all types rotate more slowly, not just stars with giant planets.
\end{abstract}

\section{Introduction}

Stars with planets (SWPs) have been shown to be more metal-rich as a group when compared to similar stars without detected planets (non-SWPs) for nearby stars, found with the Doppler method \citep{vf05}, and for stars in the {\it Kepler} field \citep{jf15}, found with the photometric transit method. SWPs and non-SWPs have also been reported to differ in vsini \citep{gg08, gg10, tak10, gg11}, but others have failed to confirm this \citep{is09,alv10,del15}.

\citet{bou08} suggested that slow rotation among SWPs caused by early star-disk interactions caused accelerated Li destruction in them. Recently, additional evidence for lower Li abundances among SWPs has been presented by \citet{fig14} and \citet{gg15}, which, if the evidence for slower rotation among SWPs is also confirmed, strengthens the Li abundance, stellar rotation and planet links. Before Bouvier's theory would be accepted as the best explanation, however, it would have to be shown that the spin-down of stellar hosts occurr very early.

The purpose of the present study is to test whether SWPs rotate more slowly non-SWPs using {\it Kepler} data and a method of analysis similar to that described in \citep{gg15}. The paper is organized as follows. In Section 2 we describe the preparation of the SWP and non-SWP comparison samples. In Section 3 we compare the measured rotation periods and simulated vsini values of SWPs and non-SWPs. We discuss the result sin Section 4 and present our conclusions in Section 5.

\section{Preparation of samples}

In order to compare the rotation periods of non-SWPs and SWPs in the {\it Kepler} field, we employed the results of the \citet{McQ14} and \citet{maz15} studies. \citet{McQ14} determined rotation periods for 34,030 main sequence targets in the {\it Kepler} field using observations from the Q3-Q14. Their dataset also excludes known eclipsing binaries and KOIs that had been identified by September 2013. In a follow-up study, \citet{maz15} applied the same analysis method as McQuillan et al. to the {\it Kepler} KOIs and published 3,356 rotation period values for the subset with detected periods. Hereafter, we will refer to these two datasets as McQ14 and Maz15, respectively. We also downloaded (on February 3, 2015) the latest list of KOIs from Q1-Q16 (hereafter, KOI16) and the stellar parameter determinations from Q1-Q17 '{\it Kepler} Stellar Table' (hereafter, S17) from the {\it Kepler} archives.\footnote{http://exoplanetarchive.ipac.caltech.edu} KOI16 includes 7,349 stars, and S17 includes 199,244 stars.

Since one of our goals is to compare stellar rotation periods in the {\it Kepler} field stars to visini values measured in nearby stars by \citet{gg11}, we must restrict the stellar parameters of the {\it Kepler} samples to the same range as the nearby stars sample. Thus, only stars with effective temperature (T$_{\rm eff}$) between 5500 and 6500 K were retained in our four {\it Kepler} datasets; exclusion of stars cooler than 5500 K also reduces the chances of including giants in the sample. In addition, only stars with log g values greater than 3.5 were retained with the express prupose of excluding giants. We removed stars with assumed solar parameters and stars lacking T$_{\rm eff}$ or log g values in the S17 dataset. This resulted in 107,633 stars. The stellar parameters from S17 replace the stellar parameters listed in the other datasets. Typical quoted uncertainties in T$_{\rm eff}$, log g, and [Fe/H] are $^{+150}_{-170}$ K, $^{+0.07}_{-0.2}$ dex, and $^{+0.20}_{-0.25}$ dex, respectively.

After removing any known KOIs from the McQ14 dataset (using the 'nkoi' column in the S17 dataset), it contained 15,467 stars. The Maz15 dataset was updated with the KOI data from KOI16. This included the disposition of each KOI: "false positive", "candidate", "not dispositioned", and "confirmed." KOIs with "false positive" disposition were removed, as were KOIs with missing rotation periods and planet size. These cuts resulted in a list of 2,332 KOIs in the Maz15 dataset.

From the equatorial coordinates listed in our revised McQ14 and Maz15 datasets, we calculated the Galactic longitude and latitude coordinates of each target. The ${\it Kepler}$ field includes target stars spread over a large range in distance from the Sun, height above the Galactic mid-plane, and interstellar extinction; however, they sample only a modest range in Galactocentric distance. In order to control for differences in Galactic location in our analysis below, we calculated the distance of each target using the following procedure.

First, we employed the 2012 version Dartmouth stellar evolution isochrones\footnote{http://stellar.dartmouth.edu/models/grid.html} for a constant age of 4 Gyrs in order to derive a simple relationship (in the form of a polynomial) between the absolute K magnitude (M$_{\rm K}$) and T$_{\rm eff}$ and [Fe/H]. Our choice of the K-band magnitude for calculating distance is motivated by the fact that it is the longest wavelength magnitude available for each target; there should be relatively little interstellar extinction in this band compared to the others. Our choice for the age is close to the mean age of simulated stars in the middle of the {\it Kepler} field, which we generated using the default settings in TRILEGAL 1.6.\footnote{http://stev.oapd.inaf.it/cgi-bin/trilegal} This relationship is possible because we are restricting our analysis to stars classified as main sequence stars. The standard deviation of the residual differences between the tabulated values of M$_{\rm K}$ from the isochrones and our simple interpolation equation is 0.14 magnitudes. Additional uncertainties in our estimates of M$_{\rm K}$ come from uncertainties in the stellar parameters; the typical uncertainty in T$_{\rm eff}$ among the {\it Kepler} targets translates into an uncertainty in M$_{\rm K}$ of 0.15 magnitudes. We calculated M$_{\rm K}$ for each star in the McQ14 and Maz15 datasets, as well as the distance, neglecting extinction.

With a distance estimate in hand for each target, it becomes possible to estimate the extinction. The total extinction along a particular sightline for a specific Galactic longitude and latitude was obtained from the NASA/IPAC online calculator.\footnote{http://irsa.ipac.caltech.edu/applications/DUST/} The \citet{sf11} reddening relation and standard $R$ value ($= A_{\rm V}/E(B-V) = 3.1$) were used. The ratio of visual to K-band extinction was set to 8.8 \citep{ind05}. Finally, the amount of extinction in each direction was assumed to reach its maximum value 3 kpc from the Sun; larger distances were not required, since the target stars are significantly distant from the midplane. The extinction for each target was scaled according to its distance relative to 3 kpc. The largest values of K band extinction were calculated to be about 0.22 magnitudes, but most values were less than 0.1 magnitudes. Nevertheless, we corrected the original distance estimates for extinction as our final step.

\section{Camparing planet hosts and non-hosts}
\subsection{Rotation Period}

In \citet{gg11} we compared vsini values of 99 SWPs and 594 non-SWPs using primarily the \citet{vf05} dataset. The SWPs among nearby stars have been detected with the Doppler method. This is in contrast to the {\it Kepler} field SWPs (classified as KOIs), which have been discovered with the photometric transit method. The nearby SWPs have accurate parallax measurements available, while very few of the {\it Kepler} field targets are close enough for parallax distance determination.

We will apply a modified version of our method of analysis described in \citet{gg08} \citet{gg10} and \citet{gg11} to the present data. In brief, in the prior studies we defined an index, called $\Delta_{\rm 1}$, which is a measure of the distance between two stars in T$_{\rm eff}$-log $g$-[Fe/H]-M$_{\rm V}$ parameter space. This index was calculated for each SWP and non-SWP and formed the basis for comparing stars from these two groups. 

We now define the following new index, $\Delta^{\rm K}_{p,c}$, which is more suitable for our present purposes:

\begin{eqnarray*}
\Delta^{K}_{p,c}=30~\vert \log~{\rm T}_{\rm eff}^{c} - \log~{\rm T}_{\rm eff}^{p} \vert +
\vert {[Fe/H]}^{c} - {[Fe/H]}^{p} \vert \\
 + 0.5~\vert \log {g}^{\rm c} - \log {g}^{\rm p} \vert + \vert {d}^{c} - 
 {d}^{p} \vert +  \vert {z}^{c} - {z}^{p} \vert
\end{eqnarray*}

where ${\it d}$ is the distance from the Sun in kpc, ${\it z}$ is the vertical distance from the mid-plane in kpc, ${\it p}$ is the KOI index, and ${\it c}$ is the comparison star (non-KOI) index. If $\Delta^{\rm K}_{p,c} = 0$ for a given pair of stars, then they are indistinguishable in these parameters.

Our decision to include the parameters ${\it d}$ and ${\it z}$ is motivated by the need to control for known Galactic-scale trends. These include the dependence of the mix of Galactic populations (thin disk, thick disk, halo) on ${\it z}$, and the concomitant trend of metallicity with ${\it z}$. In addition, thick disk stars have a different mix of $\alpha$ elements relative to Fe compared to thin disk stars; \citet{gg09} argued that differences in $\alpha$ element abundances should not be neglected when comparing SWPs and non-SWPs. For this reason, \citet{gg11} employed [M/H] in their $\Delta$ index definition. According to the TRILEGAL simulation of the $\it Kepler$ field, only about 3 \% of the main sequence stars should be thick disk stars. Therefore, their contribution should be almost negligible in our analysis. In addition, the Galactic radial metallicity gradient creates a metallicity trend with ${\it d}$, but it is a small factor for our sample. And, as noted above, extinction depends on ${\it d}$ as well. Although [Fe/H] is explicitly included as a parameter in the definition of $\Delta^{\rm K}_{p,c}$, its uncertainty is large for the typical target. However, two stars are more likely to have similar values of [Fe/H] if they have similar values of ${\it d}$ and ${\it z}$.

We employ the $\Delta^{\rm K}_{p,c}$ index in a way similar to that used to produce Figure 7 of \citet{gg15}, which is a comparison of lithium abundances in nearby SWPs and non-SWPs. In the present application, the index is used to find which non-KOIs in the McQ14 dataset are most similar to each KOI in the Maz15 dataset. The $\Delta^{\rm K}_{p,c}$ index was calculated for every pairing of a KOI and a non-KOI and then ranked.

Next, we classify each KOI star according to the size of its planet. If it is between 5 and 15 Earth radii (R$_{\rm E}$), we classify it as a ``giant.'' If it is between 2 and 5 R$_{\rm E}$, we call it a ``Neptune.'' We classify smaller planets as ``Earths.'' 

During the review cycle of the present work, {\it Kepler} Data Release 24 (DR24) was released. It is the result of the first uniform processing of the entire {\it Kepler} dataset and the first full automation of the dispositioning process. The disposition "not dispositioned" was not retained in DR24. As shown by \citep{fress13}, the rate of false positives for giant planets in the {\it Kepler} data they analyzed was about 17\%, and it was near 10\% for the other planet classes. Given the importance of eliminating as many false positives as possible and of this latest data release, we added one final step in our vetting/culling process. We cross-checked our list of KOIs in the Maz15 dataset with the DR24 dataset in order to update the dispositions. After removing the "false positive" disposition KOIs for the giant planet class, we are left with 108 KOIs with "candidate" dispositions and 43 KOIs with "confirmed" dispositons. We show the results of the comparison for the giants in Figure 1. Figure 1a shows the difference in the stellar rotation periods between each KOI star and the most similar comparison star. 

\begin{figure}
  \includegraphics[width=3.5in]{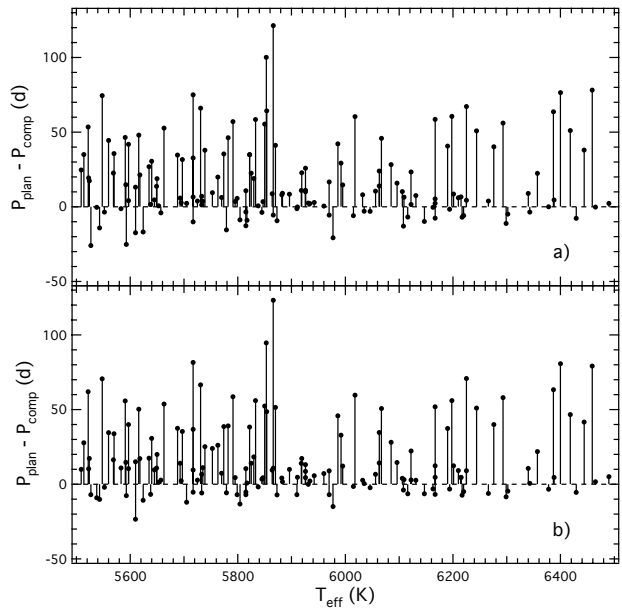}
 \caption{$\bf{a}$ Difference in the stellar rotation periods between each KOI star with a giant planet and the most similar non-KOI star. $\bf{b}$ Same as $\bf{a}$ but comparing the rotation period of each KOI star with a giant planet to the average rotation period of the three most similar non-KOI stars. See text for details.}
\end{figure}

Given the relatively large uncertainties in the stellar parameters for the {\it Kepler} targets, we added an additional step in the comparison to improve the statistics. Hence, Figure 1b is the same as Figure 1a except that the period of each KOI star is now compared to the average of the periods of the three most similar non-KOI comparison stars. Upon visual inspection, there appears to be relatively little change in Figure 1b relative to Figure 1a, except that the largest negative differences are reduced in magnitude in Figure 1b. It is very clear from both plots that the KOI stars tend to have longer rotation periods than the comparison stars. The mean differences in Figures 1a and 1b are $17.6 \pm 2.1$ days and $17.7 \pm 2.1$ days, respectively.\footnote{Here and throughout this paper, unless otherwise stated, a quoted uncertainty is the standard error of the mean.} If we compare only the 43 confirmed KOIs, the difference is only slightly reduced ($19.8 \pm 5.2$ and $18.9 \pm 4.6$ days, respectively).

We show the results for the KOIs we classify as Neptunes in Figures 2, following the same kind of analyses as those of Figures 1. Figure 2 includes 725 KOI stars with Neptunes. The mean period  differences in Figure 2a and b are both $18.0 \pm 0.9$ days. The mean period differences for the 145 confirmed stars with Neptune planets are $18.8 \pm 2.1$ and $18.4 \pm 1.9$ days, respectively.

Finally, the KOIs classified as Earths are shown in Figure 3. Figure 3 includes 878 stars. The mean period differences are $20.5 \pm 0.9$ days and $20.6 \pm 0.9$ days, respectively. The mean period differences for the 157 confirmed stars with Earths are $18.8 \pm 2.2$ and $18.1 \pm 2.1$ days, respectively.

\begin{figure}
  \includegraphics[width=3.5in]{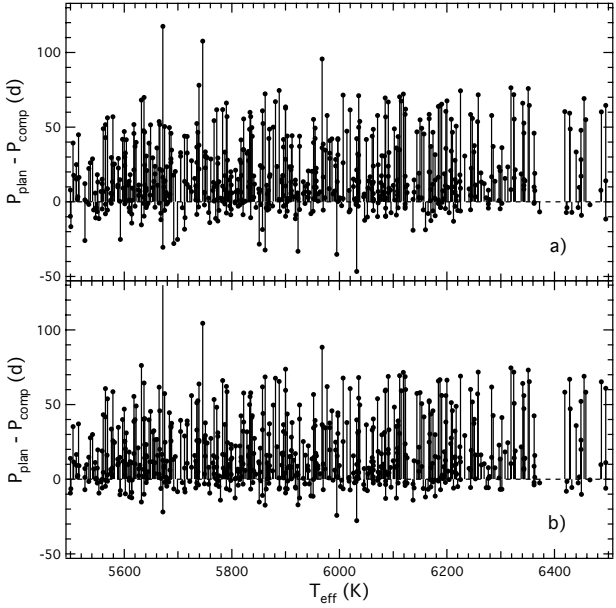}
 \caption{Same as Figure 1 but for Neptunes.}
\end{figure}

\begin{figure}
  \includegraphics[width=3.5in]{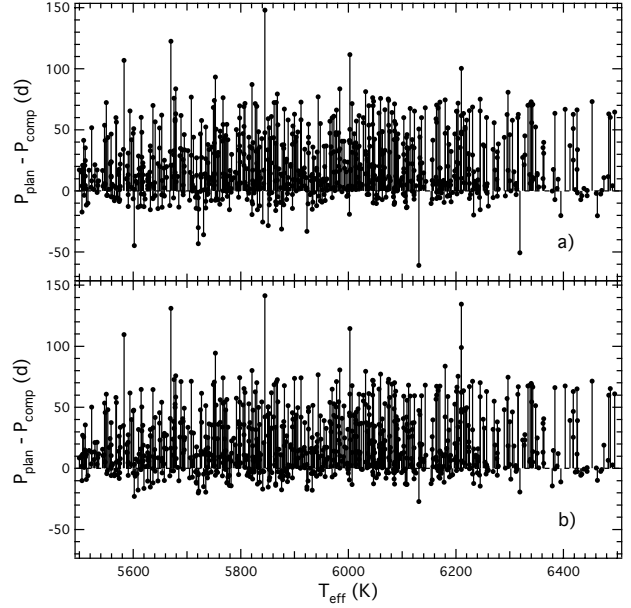}
 \caption{Same as Figure 1 but for Earths.}
\end{figure}

\subsection{vsini}

Before we can compare the results of our analysis of the stellar rotation periods of the {\it Kepler} field stars to the analysis of vsini values of nearby stars, we must account for the various differences in the way the two samples were created. To summarize, homogeneous nearby SWP and non-SWP samples, such as those described in \citet{vf05}, are based on planet detections with the Doppler method. Doppler surveys are most sensitive to massive planets with short orbital periods. Essentially all the planets of the nearby SWPs sample included in the \citet{gg11} study would fall under the ``giant'' planet classification of the present work. The non-SWPs comparison sample adopted in \citet{gg11} should contain very few giant planets. Therefore, the SWP and non-SWP samples should be relatively pure when the incidence of giant planets is being considered. When comparing rotation velocities of SWPs and non-SWPs among nearby stars in a homogeneous way, the best we can do is compare vsini values.

The {\it Kepler} field samples are very different from the nearby stars samples. Only a small fraction of planets orbiting stars in the {\it Kepler} field are detectable with the photometric transit method. The highest fraction of planet detectability occurs for the ``hot Jupiters''; about 10 \% of stars with such planets have detectable transits. This means that any non-KOI comparison sample drawn from the {\it Kepler} field targets will be contaminated with many SWPs. This is simply unavoidable with this type of survey. In the case of the giant planets, the situation is not as hopeless as it seems, since the statistically inferred true incidence of stars hosing giant planets (as we've defined them) with orbital periods $< 50$ days in the {\it Kepler} field is about 3 \% \citep{how12}. Thus, something less than 3 \% of the stars in our non-KOI comparison sample will actually host a giant planet. In our McQ14 comparison dataset (with over 15,000 stars), then, there are nearly 500 stars with undetected giant planets. Given this, it is possible that some of the comparisons in Figures 1 and 2 are between pairs of SWPs. The situation is worse for the Neptunes and Earths, as they are much more common; \citet{how12} estimate that Neptunes make up 13 \%, and \citet{pet13} find a similar fraction for Earths. However, our focus here is on giant planets.

In order to compare our results from Figure 1 to the results from \citet{gg11}, we need to convert each stellar rotation period to vsini. Our procedure is as follows. First, we adopt a probability density function (PDF) describing the distribution of the observed inclination angles of a population of stellar rotation axes; they are distributed like sini \citep{ht11}. We follow \citet{lj12} and define a PDF for sini using their equation 6 in order to produce random samples of sini. We adopt a minimum inclination angle of 5 degrees; SWPs with inclination angles smaller than this limit would be difficult to detect with the Doppler method. We produced sets of random sini values for the McQ14 and Maz15 datasets and then used them to calculate vsini for each star from its rotation period and radius. Finally, we added a fictitious error to each vsini value drawn fro a Gaussian PDF with a standard deviation of 0.5 km s$^{\rm -1}$, which is the typical uncertainty of vsini determinations in the nearby stars samples. Finally, we set any resulting vsini values that were less than 0.3 km s$^{\rm -1}$ to 0.3 km s$^{\rm -1}$ in order to be consistent with the approach of \citet{gg11}. We eliminated vsini values greater than 20 km s$^{\rm -1}$, which is necessary because planets are difficult to detect with the Doppler method around fast rotators.

We show the differences in simulated vsini between the KOI stars with giant planets and non-KOIs in Figure 4, which includes 138 stars. The mean differences in vsini in these two datasets are $-1.3 \pm 0.3$ km s$^{\rm -1}$. This is just over twice the difference quoted in \citet{gg11}, $-0.46 \pm 0.11$ km s$^{\rm -1}$, and about twice the value quoted in \citet{gg10}, $-0.66 \pm 0.13$ km s$^{\rm -1}$. The agreement is not bad, considering the very different natures of the two samples and the assumptions that went into producing the {\it Kepler} simulated vsini dataset. The difference could also be due to the fact that the {\it Kepler} KOI stars should be older, on average, than the nearby stars SWPs, given the {\it Kepler} stars' greater distance from the Galactic mid-plane.

\begin{figure}
  \includegraphics[width=3.5in]{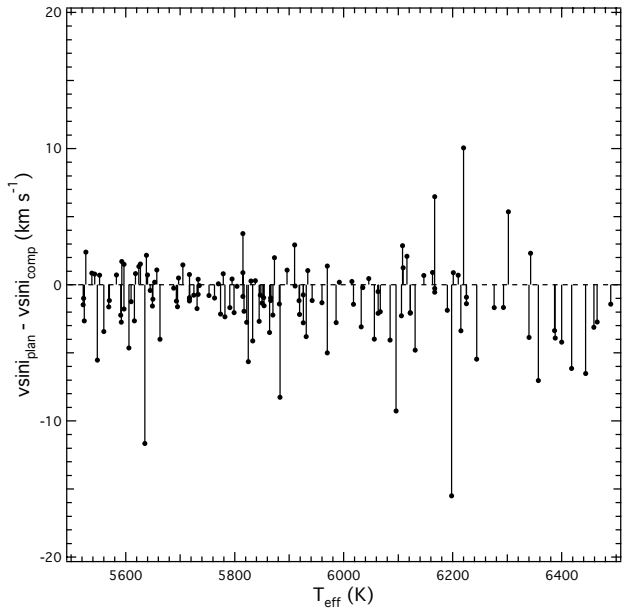}
\caption{Data from Figure 1a transformed to vsini. $\bf{b}$  See text for details.}
\end{figure}

\section{Discussion}

The results of our analysis of {\it Kepler} KOI stars confirms previous studies of nearby SWPs indicating that they rotate more slowly than stars without known planets. Our results are also consistent with the more narrow finding of \citet{McQ13} that close-in planets are rare among rapidly rotating host stars. To further explore the nature of this difference among the {\it Kepler} stars, we show in Figure 5 the rotation periods of the KOIs with giant planets and the non-KOI comparison stars. The first notable difference is evident among the slowest rotating stars. Stars in the McQ14 comparison stars dataset have a maximum rotation period of 70 days, while the KOI stars with giant planets sample have periods up to 142 days; there are 16 KOI stars with giant planets with periods longer than 70 days. Just above the ridge of greatest concentration of stars, the KOI stars have a high relative incidence. In addition, among the fastest rotators, KOI stars have a very low relative incidence.

The McQ14 and Maz15 datasets both make use of the same analysis methods on the same {\it Kepler} data (Q3-Q14). Thus, it is unlikely that the differences between the KOI and non-KOI stars evident in Figure 5 are due to differences in the data analysis. Nevertheless, as a test of the reality of the difference between the mean KOI and non-KOI stars rotation periods, we have calculated a new mean period difference for the data in Figure 1a by excluding the 16 KOI stars with giant planets with periods greater than 70 days; we obtain $11.4 \pm 1.6$ days. This is still a highly significant difference.

\begin{figure}
  \includegraphics[width=3.5in]{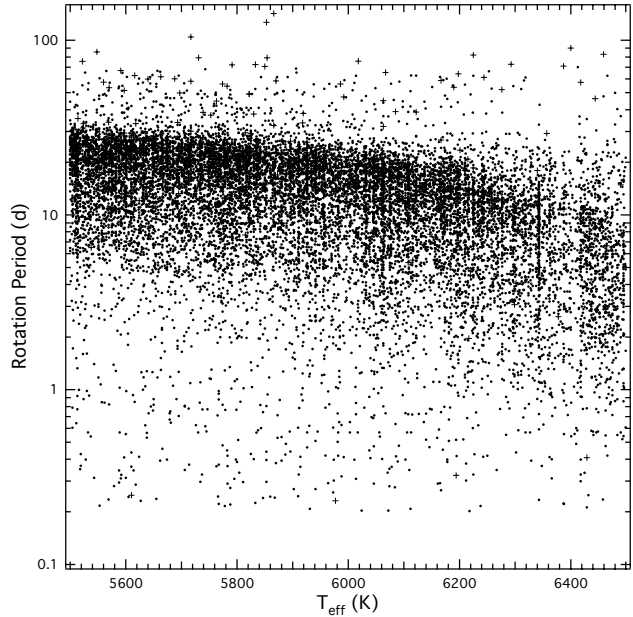}
\caption{Rotation periods of the non-KOI stars (dots) and the KOI stars with giant planets from Figure 1a (plus signs) are plotted against effective temperature.}
\end{figure}

The present results are particularly relevant to research on the lithium abundances in SWPs, which has received more attention \citep{fig14,gg15,del15}. citet{is04} suggested that since the protoplanetary disc around a young star contains a large fraction of the system’s angular momentum, it can cause rotational breaking of the star. The breaking, in turn, results in deeper mixing in the star and thus more efficient destruction of Li. \citet{bou08} placed this idea on a firmer physical foundation. The implications of this idea are important for both planet formation and stellar evolution (especially gyrochronology) theories. If correct, \citet{bou08} would place a planet-hosting star's spin-down very early in its history, while it still possessed a protoplanetary disk.

There is strong observational support for the slowing of stellar rotation with increasing age for main sequence stars younger than about 1 Gyr \citep{sod10,mei15}. Among main sequence stars in clusters with ages less than 300 Myr, the measured dispersion in rotation periods is very large, about 0.1 to 10 days. By the age of the Hyades and Praesepe ($\sim 600$ Myr), the observed dispersion in rotation period is much smaller \citep{del11}. 

Observational evidence for stellar slowdown among older stars is sparse, given that rotational modulation due to the presence of active regions becomes weaker with increasing age. This, again, is where {\it Kepler} data are particularly valuable. \citet{mei15} have measured the rotation periods of 30 faint (all fainter than $V = 15$ magnitude) main sequence stars in the 2.5 Gyr old cluster, NGC 6819. Theirs is an important study of stellar rotation, as it extends direct measurement of stellar rotation periods to ages much greater than 600 Myr. Their data show a steady trend of increasing rotation period as one goes down the main sequence with relatively small scatter. However, there is still some scatter in the periods around a mean trend curve that is greater than the quoted uncertainties in the rotation periods (see their Figures 2 and 3). In particular, the {\it Kepler} stars 5112268 and 5025271 have nearly identical colors and magnitudes and yet differ in rotation period by nearly 3 days (with period uncertainties of 0.10 and 0.25 days, respectively).

If our main result is correct, then about 30\% of the main sequence stars in NGC 6819 should be accompanied by planets and thus have longer rotation periods than stars without planets. This should result in a large scatter of the periods measured by \citet{mei15}. At first glance, the relatively small scatter found by them would seem to be inconsistent with our findings. We can offer two possible explanations for the apparent inconsistency. First, there is an observational selection bias against detecting stars with longer rotation periods. Slowly rotating main sequence stars are chromospherically less active than faster rotating stars, which makes it more difficult to measure rotation from star spots rotating into and out of view. A possible way to test this idea is to calculate the variability amplitude of main sequence stars with rotation periods 10 to 50 days longer and estimate what fraction would have been detectable in their analysis.

Second, the environment in a rich cluster like NGC 6819 might not be representative of the kind of birth environment of the typical star in the {\it Kepler} field. Only a handful of planets have been found in clusters. We need better statistics to test this idea.

\section{Conclusions}

We have verified that the rotation periods of KOI stars in the {\it Kepler} field differ significantly from those of otherwise similar non-KOI stars. Stellar rotation period measured with precision photometry is preferable to spectroscopic vsini measurements when doing this kind of comparison, given the small relative errors that can be achieved. Nevertheless, we have also shown that the rotation period measurements of KOI stars with giant planets are consistent with nearby SWP vsini measurements. Both samples show that stars with giant planets rotate more slowly than stars lacking such planets. 

\citet{bou08} suggested that slow rotation among SWPs caused by early star-disk interactions caused accelerated Li destruction in them. Recently, additional evidence for lower Li abundances among SWPs has been presented by \citet{fig14} and \citet{gg15}. These results, as well as the new findings presented in this work, further strengthen the links among Li, rotation and the presence of planets. However, additonal study is required to determine the timing of the spin-down of the host stars. Bouvier's theory requires that the spin-down would occur mostly very early in a star's history. What is surprising is that comparably slow rotation is also found among stars in the {\it Kepler} field hosting Neptunes and Earths. Also myterious within the context of these findings is the apparently small scatter in measured rotation periods of main sequence stars in old clusters.

The next steps in this line of research could involve correction for the presence of stars with planets (especially Neptunes and Earths) in the comparison samples. This kind of contamination is unavoidable in photometric transit surveys, but it should be possible to account for it in a statistical way. In addition, more rotation period measurements in star clusters are needed in order to determine if the scatter in rotation period among old stars is really as small as it appears to be from current data.

\section*{Acknowledgments}

We thank the anonymous reviewers for helpful comments.

\bsp

\label{lastpage}

\end{document}